\documentclass[12pt]{article}
\usepackage{epsf}
\newcommand{\eqnn}[1]{\begin{eqnarray*}#1\end{eqnarray*}}
\newcommand{\eqnl}[2]{\par\parbox{14.5cm}
{\begin{eqnarray*}#1\end{eqnarray*}}\hfill
\parbox{1cm}{\begin{eqnarray}\label{#2}\end{eqnarray}}}

\newcommand{\eqngrlb}[3]{\par\parbox{12.5cm}
{\begin{eqnarray}\fbox{$\displaystyle
#1\\#2$}\end{eqnarray}}\hfill
\parbox{1cm}{\begin{eqnarray}\label{#3}\end{\eqnarray}}}

\newcommand{\eqngrl}[3]{\par\parbox{14.5cm}
{\begin{eqnarray*}#1\\#2\end{eqnarray*}}\hfill
\parbox{1cm}{\begin{eqnarray}\label{#3}\end{eqnarray}}}

\newcommand{\refs}[1]{(\ref{#1})}
\def\vx{\vec{x}}
\def\intl{\int\limits}
\def\di{\displaystyle}
\def\lam{\lambda}
\def\&{&\di}
\def\bg{\begin{eqnarray}\begin{array}{rcl}\displaystyle}
\def\eg{\end{array} &\di    &\di   \end{eqnarray}}
\def\bm#1{\begin{eqnarray}\begin{array}{#1}\di} 
\def\bmo#1{\begin{eqnarray*}\begin{array}{#1}\di} 
\def\eg{\end{array} &\di    &\di   \end{eqnarray}}
\def\bgo{\begin{eqnarray*}\begin{array}{rcl}\displaystyle}
\def\ego{\end{array} &\di    &\di \nonumber  \end{eqnarray*}}

\def\btensor#1#2{\renew\left#1\begin{array}{#2}\di}
\def\etensor#1{\end{array}\right#1}

\def\ha{{1\over 2}}

\def\d{{\mbox d}}

\def\Tr{\mbox{Tr}}

\def\id{1\!\mbox{l}}

\def\ov{\over}

\def\al{\alpha}

\def\R{\mbox{l}\!\mbox{R}}

\def\CA{{\cal A}}
\def\CB{{\cal B}}

\def\CD{{\cal D}}

\def\CL{{\cal L}}

\def\CN{{\cal N}}

\def\CP{{\cal P}}
\def\CS{{\cal S}}

\def\CW{{\cal W}}

\newcommand{\mtxt}[1]{\quad\hbox{{#1}}\quad}

\date{\today}

\def\rene{\renewcommand{\arraystretch}{1.8}}
\def\renew{\renewcommand{\arraystretch}{1}}
\rene

 \newcommand{\mysection}[1]{\section{#1}\setcounter{equation}{0}}

\begin{document}

\begin{titlepage}

\parindent=12pt
\baselineskip=20pt
\textwidth 15 truecm
\vsize=23 truecm
\hoffset=0.7 truecm

\begin{flushright}
   FSUJ-TPI-98/03 \\
DIAS-STP-98/02\\
February 1998\\
 hep-th 9802191\\     
      \end{flushright}
\par
\vskip .5 truecm
\large \centerline{\bf Monopoles, Polyakov-Loops and Gauge Fixing on 
the Torus\footnote{Supported by the Deutsche Forschungsgemeinschaft,
DFG-Wi 777/3-1}} 
\par
\vskip 1 truecm
\normalsize
\begin{center}
\textbf{C.~Ford, U.~G.~Mitreuter, T.~Tok, A.~Wipf}\footnote{e--mails:
Ford, Mitreuter, Tok and Wipf@tpi.uni-jena.de}\\
\it{Theor.--Phys. Institut, Universit\"at Jena\\ 
Fr\"obelstieg 1, D--07743 Jena, Germany}\\
\vskip .5truecm
\rm\textbf{J.~M.~ Pawlowski}\footnote{e--mail: jmp@stp.dias.ie}\\
\it{ Dublin Institute for Advanced Studies,\\
10 Burlington Road, Dublin 4, Ireland}\\
\end{center}
\par
\vskip 2 truecm
\normalsize
\begin{abstract} 
We consider pure Yang Mills theory on the four torus. 
A set of non-Abelian transition functions is presented which
encompass all instanton sectors. It is argued that these transition
functions are a convenient starting point for gauge fixing. In
particular, we give an extended Abelian projection with respect
to the Polyakov loop, where $A_0$ is independent 
of time and in the Cartan subalgebra. 
In the non-perturbative sectors such gauge fixings
are necessarily singular. These singularities can be restricted to Dirac
strings joining monopole and anti-monopole like ``defects''.

\end{abstract}
\vfill
      
\end{titlepage}

\mysection{Introduction} 

A long standing and yet unsolved problem is 
to explain  color  confinement in QCD. 
An important first step in this direction would be to prove  the 
confinement of static quarks.
Indeed, this has been convincingly demonstrated by many lattice
studies \cite{lattice}.
However, it would be desirable to study this phenomenon
with less reliance on numerics, since
Monte Carlo simulations do not have the logical transparency
of mathematical derivations.
The relevant observables are 
products of Wilson-loop operators \cite{wilson}.
When one has periodicity in Euclidean time (ie.  finite temperature)
then one may  use a Polyakov loop \cite{polloops}
operator, ie. the Polyakov loop is a closed, periodic time-like 
Wilson loop.
          
Since the  Lagrangian includes massless fields and one is 
interested in the infrared behaviour of the theory, it is sensible to
implement some kind of infrared cutoff. It is well-known for example 
that in supersymmetric theories with massless modes, the
non-renormalisation
``theorems'' fail in the absence of an infrared
regulator \cite{nonren}. 
One possibility is to use a Wilsonian effective action
\cite{wilsonaction} which by 
definition includes a momentum space infrared cut-off on the quantum
fluctuations. An alternative procedure is to 
simply work on some compact Euclidean space. Of these,
the four torus, $T^4$, is the most attractive.
When dealing with $T^4$, one automatically includes the finite
temperature case ($S^1\times\R^3$), and  unphysical
curvature effects are absent. On $T^4$ one also maintains translational
invariance, and thus any relevant supersymmetry. 
              
We assume that our gauge fields are periodic in time
\bg
A_\mu(x^0+\beta,\vec{x})=A_\mu(x_0,\vec{x}).
\eg
In the quantum theory, we may interpret $\beta$ as the inverse
temperature, see for example \cite{pisarski}.
The Polyakov loop operator $P(\vec{x})$ 
is defined as the following
trace of a path ordered exponential of $A_0$
\eqnl{
P(\vec{x})=\Tr~\Gamma\left(
\CP(\beta,\vec{x})\right),
\mtxt{where}\CP(x^0,\vec{x})=\CP\exp\left[
i\int^{x^0}_0 d\tau A_0(\tau,\vec{x})\right],}{defpol}
and $\Gamma$ is the representation of the gauge group which acts on
the fermions.
The Polyakov loop is invariant under gauge transformations which are
periodic in time\footnote{
On can define a Polyakov loop operator which is  invariant
under \sl all \rm gauge transformations, ie.
$P(\vec{x})=\hbox{Tr}\left(U_0(x^0=0,\vec{x}){\CP}(\beta,\vec{x})\right)$,
$U_0(x^0,\vec{x})$ being the time transition function (see section 2).
However, since we always work with time periodic objects, ie.
$U_0=\id$, definition \refs{defpol} is sufficient for our purposes.}. 
The two-point function
\eqnl{
e^{-\beta F(\vx ,\vec y)}= 
\langle P(\vx ) P^\dagger(\vec y)\rangle_{\beta}}{freeenergy}
yields the free energy $F(\vx ,\vec y)$ in the presence of a heavy
quark  at $\vx $
and a heavy antiquark at $\vec y$.  In the confining 
low-temperature phase $F(\vx ,\vec y)$ increases for 
large separations\footnote{We assume that the three spatial 
edge lengths of our torus are much
larger than $\Lambda_{QCD}^{-1}$}
of the  quark-antiquark pair and thus
$\langle P(\vx )P^\dagger (\vec y)\rangle\to 0$.
In the deconfining high-temperature phase the free energy 
reaches a constant value for large separations and
$\langle P(\vx )P^\dagger (\vec y)\rangle\to \hbox{constant}\neq 0$.
Inferring the cluster property we see that
$\langle P\rangle_\beta$ vanishes in the confining
phase but not in the deconfining one. In other words,
it is an order parameter for confinement. 
                    
Note  that the Weyl gauge, $A_0=0$, is not compatible with 
time-periodicity. 
Yet we still would like $A_0$ to be as simple as possible, since we
are interested in observables only depending on $A_0$. 
On the two dimensional torus,
$T^2$, one can perform an Abelian projection \cite{tHooft} with respect
to the Polyakov loop operators and gauge fix in such a way
that $A_0$ is in the Cartan subalgebra and is independent of
time, while preserving the time periodicity of $A_1$.
In this gauge one has a remarkable cancellation between
part of $\exp(-S)$ and the Fadeev-Popov determinant.
This simplifies the calculation of the partition function
and the expectation value of the Polyakov loop order parameter, 
and avoids  zero mode ambiguities \cite{mpw}.
               
In this paper we address the question of to what extent
the gauge fixing used in \cite{mpw} can be generalised to 
QCD on the four torus.
The gauge fixing procedure hinges on the  diagonalisation
of the path ordered exponential, ${\cal P}(\beta,\vec{x})$,
 whose trace is the Polyakov loop.
It is convenient that ${\cal P}(\beta,\vec{x})$
be periodic in the spatial variables.
Yet unless we are in the perturbative sector, the gauge fields
themselves are necessarily non-periodic.  
This non periodicity is characterised by a set of group-valued transition
functions.
We introduce a set of \textit{ non-Abelian}  transition functions which
facilitate a periodic ${\cal P}(\beta,\vec{x})$  even in the non-perturbative
sectors.
Unlike the well known Abelian transition functions introduced by
't Hooft \cite{thooft}, our transition
functions encompass all instanton sectors, thus solving the
problem of finding smooth transition functions for the odd instanton 
sectors of $SU(2)$ gauge theory.

In contrast to the two dimensional case the 
diagonalisation procedure has unavoidable  singularities. The
singularities can be interpreted as Dirac strings \cite{dirac} joining 
magnetically charged  ``defects''. 
Here we understand defects as points, loops  (not to be confused with
the
Dirac strings!), sheets and lumps where ${\cal P}(\beta,\vec{x})$
has degenerate eigenvalues.
The locations of the defects are gauge invariant 
 and may be viewed as  additional 
``collective coordinates'' associated to the gauge fixing.
The simplest case (and probably the most relevant for the QCD
path integral) is where one only has point 
defects (which can be viewed as magnetic monopoles).
 Here the final gauge fixed potential has very simple
periodicity
properties,
and the topological charge is completely fixed by the network of
monopoles and Dirac
strings (see also \cite{reinhardt}).
We also consider the more general case where 
one has extended defects.

The outline of this paper is as follows. In section two we
recall some basic facts about gauge fields on the torus, including
the standard Abelian transition functions. Our new set 
of non-Abelian transition functions (which includes the odd
sector of $SU(2)$) is presented in section 3. We also explain how
one can use the Polyakov loop itself to define a different  
set of non-Abelian transition functions. Next, in section 4 
we elaborate  our gauge fixing procedure, and study the special case
where one has no defects.
In section 5 we
discuss the problem of defects, and show how they contribute to the
instanton number. 
Conventions and 
some technical results are collected in three appendices.

\mysection{Gauge fields on $T^4$}

We view the four torus as $\R^4$ modulo the lattice generated by
four orthogonal vectors $b_\mu,\quad \mu=0,1,2,3$.
The Euclidean lengths of the $b_\mu$ are denoted by
$L_\mu$ (we may identify $L_0$ with the inverse temperature $\beta$).
 Local gauge invariants such as $\hbox{Tr}F_{\mu \nu} F_{\mu \nu}$
are periodic with respect to a shift by an arbitrary lattice vector. 
However, it follows that gauge fields have to be periodic only up to gauge 
transformations. In order to specify boundary conditions for gauge potentials
$A_\alpha$ on the torus one introduces a set of \textit{  transition
functions }
$U_\mu(x)$, which are defined on the whole of $\R^4$.
The periodicity properties of $A_\alpha$ are as follows
\bg
A_\alpha(x+b_\mu)
=U_\mu^{-1}(x)A_\alpha(x)U_\mu(x)+i
U_\mu^{-1}(x)\partial_\alpha U_\mu(x),
\quad \mu=0,1,2,3\eg
where the summation convention is \textit{ not  } applied.
The transition functions 
$U_\mu(x)$ have to satisfy the  cocycle condition\footnote{
One can consider the more general possibility
$U_{\mu}(x)U_{\nu}(x+b_\mu)=Z_{\mu\nu}U_\nu(x)U_\mu(x+b_\nu)$
where the \sl twists \rm $Z_{\mu\nu}$ lie in the centre of the group.
In this paper we concentrate on the untwisted case, ie.
$Z_{\mu\nu}=\id$,
which is appropriate if the matter fields are in a fundamental
representation of the gauge group.}
\eqnl{
U_{\mu}(x)U_{\nu}(x+b_\mu)=U_\nu(x)U_\mu(x+b_\nu).}{cocycle}
Under a gauge transformation, $V(x)$, the pair $(A,U)$ is mapped to 
\eqnl{ A_\alpha^V(x)=V^{-1}(x)A_\alpha(x)V(x)+i
V^{-1}(x)\partial_\alpha V(x),\ \ \ 
U_\mu^V(x)=V^{-1}(x)U_\mu(x)V(x+b_\mu).}{newU}
We define the (integer valued) 
topological charge or instanton number as follows 
\eqnl{
q={1\ov 32\pi^2}\int_{T^4}\epsilon_{\mu\nu\al\beta}\hbox{Tr}\,F_{\mu\nu} 
F_{\al\beta}.}{topological}
The integrand in \refs{topological} can be written as a total
derivative. Using Stokes theorem we get 
\eqngrl{
q &=& \frac{1}{24\pi^2}\sum_\mu \int_{\CB_\mu} 
\epsilon_{\mu\nu\rho\sigma}\Tr \left[(U^{-1}_\mu\partial_\nu U_\mu) 
(U^{-1}_\mu\partial_\rho U_\mu)( U^{-1}_\mu\partial_\sigma U_\mu)\right]}
{&& -\frac{1}{8\pi^2} \sum_{\mu,\nu} \int_{\CB_{\mu\nu}}\epsilon_{\mu\nu\rho\sigma}
\hbox{Tr} \left[( U^{-1}_\nu\partial_\rho U_\nu)
(\partial_\sigma U_\mu(x+b_\nu)U^{-1}_\mu(x+b_\nu)) \right],}{q}
with
\eqnn{
\CB_\mu = \{x\in T^4|x_\mu=0\},\qquad \CB_{\mu\nu}=
\{x\in T^4|x_\mu=x_\nu=0\}.}
(see also \cite{vanbaal}).
That is $q$ is fully determined by the transition functions.
In particular, if we take all the transition functions to be the
identity (i.e. we assume the gauge fields are periodic in all directions)
then the instanton number is zero. Accordingly, if we are to 
describe the non-perturbative  sectors, one must consider 
non-trivial transition functions.
For a given $q$ we only require \textit{ one } set of transition functions.
If we have two sets of transition functions with the same
instanton  number then they are gauge equivalent \cite{vanbaal}.

For $SU(N)$, $N>2$, one can write down a set of
very simple \textit{ Abelian } transition functions, which
include all possible values of $q$.
For $SU(2)$, the situation is rather peculiar, in that there exist
Abelian transition functions for the even instanton number
case, but for   odd $q$, the transition functions
are \textit{  necessarily } non-Abelian.

Consider the following set of transition functions
\eqnl{
U_0=U_2=\id,\quad
U_1(x)=e^{2\pi i H_1 \xi_2},\quad
U_3(x)=e^{2\pi i H_3 \xi_0},}{trans}
where  $H_1,H_3\in \CL$, 
with $\CL$ being  the discrete lattice in the
Cartan subalgebra ${\cal H}$;
\eqnl{
\CL\equiv \left\{ H\in {\cal H}|e^{2\pi iH}=\id\right\}}{discrete}
and we have introduced the dimensionless coordinates
\bg\label{dim}\xi_\mu=x_\mu/L_\mu,\quad \mu=0,1,2,3.\eg
These transition functions satisfy the cocycle condition (\ref{cocycle}),
and using \refs{q}, the instanton number  associated with these transition functions
is simply
\eqnl{
q=\Tr \,H_1 H_3.}{charge-abelsch}
Now, if we take $H_3$ to be proportional to $H_1$
it is easy to see that $q$ is always even. To get an odd charge
one must take non-parallel $H$'s. For example, in $SU(3)$ consider
\eqnn{
H_1=\btensor{(}{ccc}
1&0&0\\0&-1&0\\0&0&0\etensor{)}\quad
\hbox{ and }\quad  
H_3=\btensor{(}{ccc}0&0&0\\0&-1&0\\0&0&1\etensor{)}.}
In this case $q=1$.
However, for $SU(2)$ $H_1$ and $H_3$ must be parallel since
the Cartan subalgebra is one dimensional. Hence, within this class of
transition functions one is restricted to even topological charges. 
Although the transition functions \refs{trans} are not the most general
Abelian transition functions, it is easy to see that \textit{ any } 
Abelian transition functions lead to an even $q$ for 
$SU(2)$.

Although we have concentrated on the transition function question,
another (to date unsolved) problem is to obtain the instantons for pure
gauge theory on $T^4$. While 't Hooft found some extremely simple
``Abelian'' instantons \cite{thooft}, these can only represent single
points in the moduli space of a given instanton sector. 
This is in sharp contrast to the situation on $S^4$ where
Atiyah \textit{ et al  } \cite{Atiyah}  gave an algebraic recipe for computing all
instantons.
In fact one of the few things known about instantons on $T^4$
is a negative result. Using the Nahm transformation \cite{nahm},
van Baal \cite{vb} 
has argued that there are no $SU(N)$ instantons with 
 $q=1$\footnote{However, by using the Nahm-transformation 
one can construct transition functions and instanton
solutions with $q=1$ for $U(N\geq 1)$}.
We should stress that  while there are \textit{no charge one instantons}
there do exist \textit{ configurations } with $q=1$.
While for $q=1$, the minimal action is never achieved
one can find configurations whose action is arbitrarily close
to the instanton number.
Numerical \cite{perez} and analytical studies indicate that as one
brings the action closer to the minimum the action density
becomes concentrated near a point.
For the higher charge $|q|>1$ sectors,
smooth instantons are known to exist \cite{someone}.
However, for the purposes of our gauge fixing the explicit
form of the instantons is not required.

\mysection{Non Abelian transition functions and Polyakov loops}

We have seen that Abelian transition functions are not sufficient
to describe the odd charge sectors, for the gauge group
$SU(2)$. Yet we would still like to have our transition functions
as simple as possible. Consider the following possibility;
let us take three of the four transition functions to be the identity, ie.
\eqnl{
U_0=U_1=U_2=\id .}{idea}
Within this ansatz, the cocycle condition \refs{cocycle} implies
that $U_3(x)$ is periodic in $x_0$, $x_1$ and $x_2$.
Now the formula \refs{q} for the instanton number reduces to
\eqnl{
q(U_3)=\frac{1}{24\pi^2}\int_{\CB_3} 
\epsilon_{3\nu\rho\sigma}\Tr\left[ (U^{-1}_3\partial_\nu U_3 
)(U^{-1}_3\partial_\rho U_3)( U^{-1}_3\partial_\sigma U_3)\right]}
{3dintegral}
with $\CB_3=\{x\in T^4|x_3=0\}$. 
Note that the two dimensional integrals in \refs{q} drop out, 
and one only has a single three dimensional integral. 
Furthermore, it is evident that the
$x_3$ dependence of $U_3$ is irrelevant, and we may assume that
$U_3$ is independent of $x_3$. In other words, suppose we have
a $U_3(x)$ which depends on $x_3$, then a simpler $U_3$ with
the same instanton number can be obtained simply by setting $x_3$ to be
an arbitrary constant. A very useful consequence of 
\refs{3dintegral} is that if $U_3(x)$ can be decomposed into
\textit{ periodic } factors, then the topological charge is simply a sum
of the contributions of the periodic factors, more precisely,
if we can write $U_3(x)=P_1(x)P_2(x)$, where $P_1(x)$ and $P_2(x)$ 
are periodic in all directions, then
\eqnn{
q(U_3)=q(P_1P_2)=q(P_1)+q(P_2),}
much like the situation on $S^4$. 

First we show that \refs{idea} is 
easily achieved in the even sectors of $SU(2)$.
Let us start with \textit{ Abelian } transition functions 
\eqnl{
U_0=U_2=\id,\quad U_1=e^{2\pi i\xi_2\sigma_3},\quad
U_3=e^{2n\pi i \xi_0\sigma_3}}{2n}
which lead to $q=2n$, $n\in Z$ (we use the dimensionless coordinates
$\xi_\mu$ defined by \refs{dim}). Here only two transition functions are the 
identity. However it is straightforward to gauge transform $U_1$ to unity.
To achieve this we require a gauge transformation $V(x)$ which
is periodic in $x_0$ and $x_2$ (since we wish to keep $U_0$ and $U_2$
as unity), and has the property that
\eqnn{
V(x+b_1)=e^{-2\pi i \xi_2\sigma_3}V(x).}
Choosing the  parameterisation 
\eqnl{
V(x)=\btensor{(}{cc} \al(x)&  \beta^*(x)\\-\beta(x) & \al^*(x)\etensor{)},
\quad |\al|^2+|\beta|^2=1,}{V}
$\al(x)$ and $\beta(x)$ are periodic in $x_0$ and $x_2$, and satisfy
\eqnn{
\al(x+b_1)=e^{-2\pi i \xi_2}\al(x),\quad
\beta(x+b_1)= e^{2\pi i \xi_2}\beta(x).}
One can simply take (our conventions regarding theta functions are
explained in Appendix A)
\eqnl{
\al^*(x)=
{1\over \CN}\theta
\btensor{[}{c}\xi_2 \\\di \xi_1\etensor{]}(0,i),\quad
\beta(x)={1\over \CN}\theta
\btensor{[}{c}\xi_2 \\\di \xi_1\!+\!d\etensor{]}(0,i),
\qquad |\alpha|^2+|\beta|^2=1,}
{gaugetrans}
where $d$ is not an integer.
Since the two theta functions  are regular  and have no common zeroes, 
the functions $\al(x)$ and $\beta(x)$ are smooth.
Note that $V(x)$ only depends on $x_1$ and $x_2$.
After this gauge transformation \refs{idea} holds and
$U_3(x)$ becomes non-Abelian
\eqnn{
U_3=V^{-1}(x_1,x_2)e^{2\pi i n \xi_0 \sigma_3}V(x_1,x_2).}
Multiplying $U_3(x)$ by the periodic Abelian factor $e^{-2\pi i n \xi_0
\sigma_3}$ does not change the instanton number.  Hence, an equally 
valid set of transition functions for the $2n$  sector is
\eqnn{
U_0=U_1=U_2=\id, \qquad U_3=V^{-1}(x)\,e^{2\pi i n\xi_0\sigma_3}\,V(x)\,
e^{-2\pi i n \xi_0\sigma_3}.}
Note that $U_3$ is independent of $x_3$ and periodic in $x_0$, $x_1$
and $x_2$. Now consider the following set of transition functions
\bg
U_0=U_1=U_2=\id,\qquad U_3=V^{-1}(x)e^{\pi i n  \xi_0 \sigma_3}
V(x)e^{-\pi i n \xi_0\sigma_3},\quad n\in Z.\eg
 $U_3(x)$ is still periodic in $x_0$, $x_1$ and $x_2$ and thus these
transition functions satisfy the cocycle condition (\ref{cocycle}).
  It is easy to see
that the instanton number 
of these transition functions is  precisely 
half  that of \refs{2n}; i.e. now we have $q=n$, $n\in Z$.
Thus we have a set of $C^\infty$ transition functions for all instanton 
sectors. Let us write our $U_3$ more explicitly
\eqnn{
U_3(x)=\btensor{(}{cc}
 |\al|^2+|\beta|^2e^{-2\pi i n \xi_0}&
\al^* \beta^* (e^{2\pi i n \xi_0}-1)\\ 
\al\beta(1-e^{-2\pi i n \xi_0})& |\al|^2+|\beta|^2
e^{2\pi i n \xi_0}\etensor{)}.}
Note that 
\eqnl{
U_3(x_0=0,\vx )=\id.}{prop}
This will greatly simplify the analysis of the gauge fixing in the
next section. 

Suppose we have a set of transition functions with the following properties
\eqnl{
U_0=\id,\ \ U_i(x_0=0,\vx )=\id.\quad i=1,2,3.}{condition}
The non-Abelian transition functions introduced here clearly satisfy
these conditions. 
Then  consider the following gauge transformation
\eqnn{
V(x_0,\vx )=\CP (x_0,\vx ),}
where $\CP (x_0,\vx )$ is the path ordered exponential
in \refs{defpol} which in general is non-periodic in time.
Now $\CP (x_0,\vx )$ has the following periodicity properties
\eqngrl{
\CP (x_0+L_0,\vx )&=&\CP(x_0,\vx )\CP (L_0,\vx )
}
{\CP (x_0,\vx +b_i)&=&U_i^{-1}(x_0,\vx )\CP (x_0,\vx )
U_i(x_0=0,\vx ).}{polloopperiod}
For brevity we use the notation
 \bg\CP(\vx):=\CP(L_0,\vx).\eg
Using (\ref{newU},\ref{condition},\ref{polloopperiod}), the gauge transformed 
transition functions are
\eqnn{
U_0^V=\CP(\vx ),\qquad U_1^V=U_2^V=U_3^V=\id.}
Thus we have performed a gauge transformation from transition
functions where $U_0=U_1=U_2=\id$ to transition functions
with $U_1=U_2=U_3=\id$.
Note however that the new $U_0$ is simply the path ordered
exponential of the original $A_0$ whose trace is the Polyakov
loop. 
Applying the formula \refs{q} for the instanton number
to the new set of transition functions yields
\eqnl{
q=\frac{1}{24\pi^2}\intl_{\CB_0}\epsilon_{0 i j k}\hbox{Tr}\left[
(\CP^{-1}\partial_i \CP)(\CP^{-1}\partial_j \CP)(\CP^{-1}\partial_k
\CP)\right],}{polloopindex}
where $\CP=\CP(\vx )$, and ${\cal B}_0=\{x\in T^4|x_0=0\}=T^3$.
It is evident that for $SU(2)$ the right hand side is the
 winding number of the map ${\cal P} :T^3\rightarrow SU(2)\cong S^3$,
 ie. the instanton number is  just the winding number of the Polyakov loop.
 The analogous result for gauge theories on $\R^4$ has been given
 in ref. \cite{reinhardt}.         
We emphasise that \refs{polloopindex} is only 
valid when the (original) transition functions satisfy \refs{condition}.

\mysection{Gauge fixing  on $T^4$ without Defects}

We may always assume that we start with a smooth gauge potential 
which is periodic in time, so that $U_0=\id$. 
We may also assume that we are in a gauge where the spatial transition
functions have the property \refs{condition}.
Thus we may use the formula \refs{polloopindex} for the
instanton number.
Another useful consequence of (\ref{condition}) is that this together
with
(\ref{polloopperiod}) implies that ${\cal P}(\vx)={\cal P}(L_0,\vx)$
is \textit{  periodic  in all spatial directions. }
Note that the standard Abelian transition functions can only have 
property (\ref{condition}) if we are in the perturbative ($q=0$) sector.
The non-Abelian transition functions given in the last section do
indeed satisfy (\ref{condition}).

Following \cite{mpw} we seek a (time-periodic) gauge transformation, $V(x)$,
for which the gauge transformed $A_0$ is independent of time and in
the Cartan subalgebra.
Below we argue that it is impossible in general 
to find a smooth gauge transformation
which leads to a gauge field with the desired properties.
While it is straightforward to \textit{  formally }
define a suitable gauge transformation, the gauge transformed
potential
is ill defined in the presence of ``defects'' where
${\cal P}(\vx)$ has degenerate eigenvalues \cite{tHooft,reinhardt}.
This motivates us to define the \textit{ defect manifold }
\eqnl{
\CD=\{\vx\in T^3\vert \CP(\vx)\mtxt{ has at least one degenerate
eigenvalue.\}}}{defectmani}
which is invariant under time-periodic gauge transformations.
A \textit{  defect } is understood  to be a connected subset of ${\cal D}$.

Before we consider the various defects, we first show that in the
absence of
defects a suitable non-singular gauge transform exists.
More precisely, 
 for $\CD=\emptyset$,
there is a smooth (periodic in time but non-periodic in the spatial 
variables) gauge transformation which transforms our starting gauge field, 
so that $A_0$ has the simple form
\eqnl{
A_0=a_0(\vx )}{eichfix}
with $a_0(\vx )$ in the Cartan subalgebra and periodic,
\eqnl{
a_0(\vx+b_i)=a_0(\vx),\qquad i=1,2,3.}{periodic}
Consider the time-periodic gauge transformation  \cite{mpw}  
\eqnl{
V(x_0,\vx )=\CP(x_0,\vx ) \CP^{-\xi_0}( \vx)\,W(\vx ),}
{gauge-trf-1}
where $\CP(x_0,\vx )$ is the path ordered exponential \refs{defpol},
and $W(\vx)$ diagonalises $\CP(\vx)$, i.e.
\eqnl{
\CP(\vx)=W(\vx) D(\vx) W^{-1}(\vx ),\qquad
D(\vx)=\exp\{2\pi i H(\vx )\},}{diagonalization}
with $H(\vx )$ in the Cartan subalgebra ${\cal H}$.
The fractional power of $\CP$ in \refs{gauge-trf-1} is defined
via this diagonalisation of $\CP$.
Then it follows at once 
that the gauge transformed $A_0$ reads
\eqnl{
{A}_0^V=\frac{2\pi}{L_0}H(\vx ).}{a0}

\begin{figure}[ht]
\begin{minipage}[ht]{16cm}
\centerline{\epsfysize=5 cm\epsffile{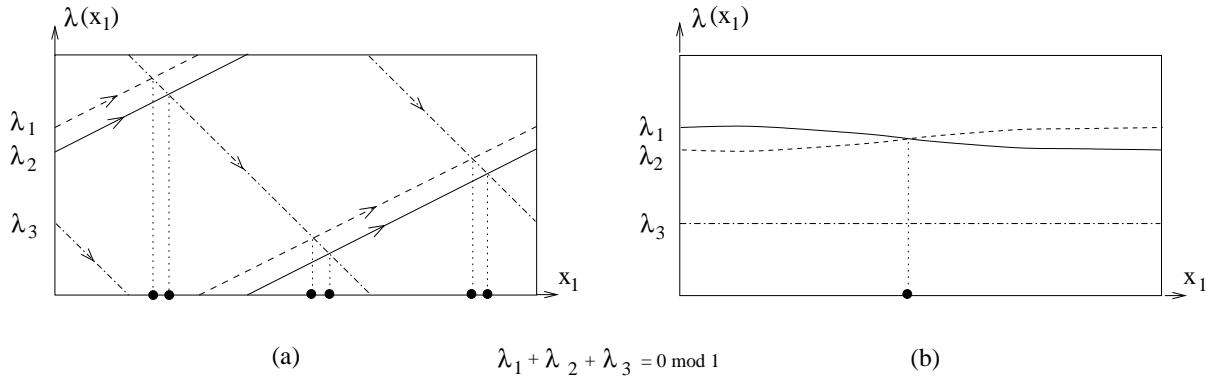}}
\caption{\label{spectorus}\textsl{The eigenvalues $\lam_i$ of $H(\vx)$
may wind (a) or there may be a spectral flow (b). In
both cases $\CP$ has degenerate eigenvalues
at $\bullet$. Shown are examples for $SU(3)$.}}
\end{minipage}
\end{figure}

 For $\CD=\emptyset$
the eigenvalues of $\CP$ are nowhere degenerate and we
can find smooth $D(\vx),W(\vx)$. 
Since  $\CP(\vx)$ {is periodic in all spatial directions}
it has  the same spectrum at $\vx$ and
$\vx+b_i$. In the absence of defects the spectral flow 
from $\vx$ to $\vx+b_i$ cannot interchange two eigenvalues,
that is the situation  depicted in fig.\ref{spectorus}b
cannot occur, and $D(\vx)$ must be periodic.
In general, the periodicity of $D(\vx)$ implies only that the eigenvalues
of $H$ are periodic modulo $1$. But if they are not
periodic they would have to wind as shown in 
fig.\ref{spectorus}a when we move from
$\vx$ to $\vx+b_i$. Then at least 
one eigenvalue of $H$ is degenerate somewhere on $T^3$ and $\CD$
is not empty. Thus $H(\vx)$ must be periodic,
\eqnl{H(\vx +b_i)=H(\vx ).}{hilfs}
>From \refs{a0} it is clear that the transformed $A_0$ indeed has the 
stated properties.

If $\CP$ has no degenerate eigenvalue, then $W$
in \refs{diagonalization} and hence the gauge
transformation $V$ in \refs{gauge-trf-1} is determined up to 
right-multiplication by a diagonal matrix.
The role of the residual local gauge group 
$U(1)^{N-1}\subset SU(N)$,
and in particular the transformation properties
of the various matter and gauge fields under this
residual symmetry, has been discussed lucidly in \cite{tHooft,schierholz}.

Hence, although $\CP(\vx)$ and $\exp(2\pi iH(\vx))$ are periodic, 
we may not assume that $W(\vx)$ is periodic. All we can say is that
\eqnl{
W(\vx+b_i)=W(\vx )R_i(\vx),\quad i=1,2,3,}{hilfs1}
where the
$R_i(\vx)$ 
lie in the residual gauge group $U(1)^{N-1}$, i.e. they are 
Abelian and satisfy the cocycle condition
\bg\label{cocycle-II}
R_i(\vx )R_j(\vx +b_i)=R_j(\vx )R_i(\vx +b_j).\eg
Using (\ref{newU},\ref{polloopperiod}) and \refs{condition}, 
the final form of the transition functions is $U_0^V=\id$ and
\eqnl{
U_i^V&=&e^{2\pi i \xi_0\, H(\vx )}
W^{-1}(\vx )W(\vx +b_i)e^{-2\pi i \xi_0\, H(\vx )}=R_i(\vx ),}
{transition-fct-2D}
which are Abelian. Inserting these
transition functions into \refs{q} yields $q=0$.
This already shows, that for gauge fields with
non-zero instanton number there are necessarily defects on $T^3$.

In the $SU(2)$ case we may write
\eqnn{
R_i(\vx)=e^{2\pi i r_i(\vx) \sigma_3},}
where the $r_i(\vx)$ are functions of the spatial coordinates.
The cocycle condition (\ref{cocycle-II}) implies that
\eqnl{
 \epsilon_{kij}\Big(r_i(\vx+b_j)-r_i(\vx)\Big)=n_k\in Z.}{integers}
Unless all the $n_k$ are zero one cannot find a smooth 
diagonalising $W(\vx)$ such that the Abelian 
transition functions $R_i(\vx)$ become the identity. 
We have seen, that $W(\vx)$, which diagonalises $\CP(\vx)$
is defined only up to right-multiplication,
\eqnl{
W(\vx)\longrightarrow W(\vx)\;e^{i\lam(\vx)\sigma_3}.}{resgauge}
If we append to each point in $T^3$ the set of all
diagonalising matrices $W(\vx)$, we get a
$U(1)$-principal bundle over $T^3$, here denoted by
$Q(T^3,U(1))$ \cite{griesshammer}. 
A smooth and \textit{periodic} $W(\vx)$ on $T^3$
would be a global section in this bundle. But in
general the $U(1)$-bundles over $T^3$ are non-trivial
and are characterised by three integers.
Indeed, with a (time-independent) Abelian gauge transformation 
\refs{resgauge} we can bring the transition functions $R_i$ into 
the standard form
\eqnl{
R_1=\id, \qquad R_2=e^{-2 \pi i n_3 \xi_1\sigma_3}\mtxt{and}
R_3=e^{2 \pi i ( n_2 \xi_1 - n_1 \xi_2 )\sigma_3},}{simpletrans}
where the $n_i$ are the integers
defined in \refs{integers}. If not all $n_i$ vanish, then
these are transition functions of nontrivial $U(1)$-bundles 
over $T^3$.

A more direct and physical way to understand the obstruction uses
the (magnetic) $U(1)$-gauge potential \cite{tHooft,schierholz}
\eqnl{
A_{\rm mag}={1\ov 2i}\Tr \Big(W(\vx )^{-1}\d W (\vx )\sigma_3\Big)}{magnpot}
on $T^3$, which transforms under the residual gauge transformation 
\refs{resgauge} as
\eqnn{
A_{\rm mag}\longrightarrow A_{\rm mag}+\d \lambda.}
Using \refs{cocycle-II} it follows at once that the $3$ magnetic fluxes
\eqnn{
\Phi_i=\intl_{x_i=const}\!\!\!\!F_{\rm mag}=
\intl_{x_i=const}\!\!\!\!dA_{\rm mag}
= \epsilon_{ijk}\Big(r_j(x+b_k)-r_j(x)\Big)=2\pi n_i}
are quantised.
We conclude that the integers $n_i$ in (\ref{integers},\ref{simpletrans}) 
cannot be changed by a smooth (Abelian) gauge transformation. 
Note, that the flux $\Phi_i$ is independent of $x^i$
and $A_{\rm mag}$ may be interpreted as a sourceless magnetic potential
permeating  the torus.

The fixing of the residual gauge freedom can be accomplished
much like in the two-dimensional case \cite{mpw} and is discussed
in appendix C.

\mysection{Gauge fixing with defects}

\begin{figure}[ht]
\begin{minipage}[ht]{15cm}
\centerline{\epsfysize=14 cm\epsffile{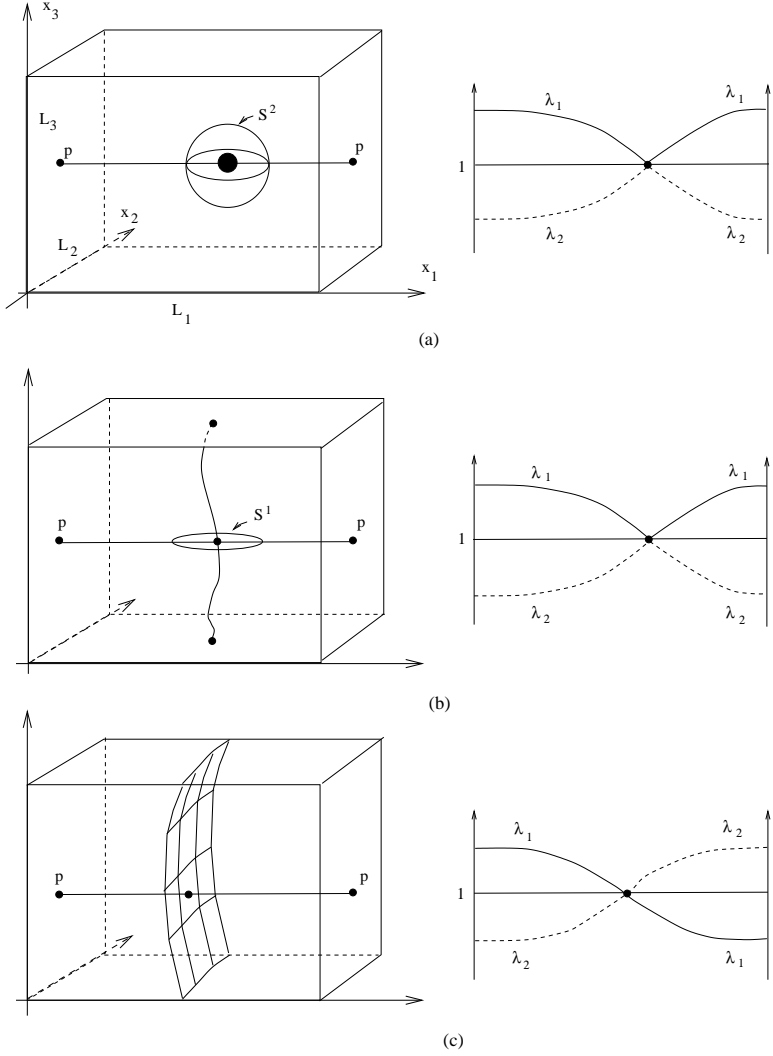}}
\caption{\label{defects}\textsl{Monopoles, loops, domain walls and
spectral flows}}
\end{minipage}
\end{figure}

Below we shall argue that isolated defects may be identified with
magnetic monopoles, line defects  with magnetic loops and
sheetlike defects with domain walls.
Monopoles may be present if ${\cal D}^c=T^3\setminus{\cal D}$ contains
non-contractable 2-spheres and magnetic loops if
${\cal D}^c$ has non-contractable loops (besides the 3 topologically
distinct
loops winding around $T^3$).
In other words, monopoles and loops 
can only be present if
\begin{center}
\begin{tabular}{l|l}\hline
$\pi_2(\CD^c)\neq 0$ & monopoles\\
$\pi_1(\CD^c)\neq Z^3$ & loops\\\hline
\end{tabular}
\end{center}
Besides monopoles and loops, there may exist
 defect walls extending over the whole three torus. The three
types of defects are depicted in fig.\ref{defects}.

We could try to repeat the analysis of chapter 4  in the
presence of defects still assuming
that $W$ and $D$ in \refs{diagonalization} are smooth.
If only monopoles and loops are present, then we can
connect $\vx$ with $\vx+b_i$ by a path in $D^c$.
Along such a path the eigenvalues of $H$ can neither wind
nor exchange  as in the absence of defects. 
Hence, spectral flows as shown in fig.\ref{spectorus}
are not possible and $H(\vx)$ must be periodic, and as in chapter 4
we have
$U^V_i=R_i(\vx)$. Since such transition functions have instanton number
zero,
  we have a contradiction in all $q\neq 0$
sectors.

\begin{figure}[ht]
\begin{minipage}[ht]{15cm}
\centerline{\epsfysize=7 cm\epsffile{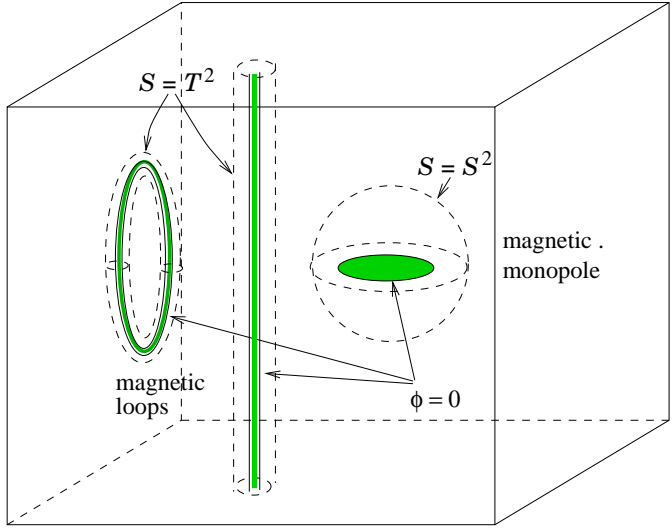}}
\caption{\label{winding}\textsl{The magnetic flux
through closed surfaces is equals the winding of
the normalised Higgs field $\CS\to S^2$}}
\end{minipage}
\end{figure}

We now specialise to the gauge group $SU(2)$; we will consider $SU(N)$
elsewhere.
The defect manifold is now simply
\eqnn{
{\cal D}=\{\vx\in T^3| {\cal P}(\vx)=\pm \id\}.}
Thus we have two distinct defect sets,
according to whether ${\cal P}(\vx)$ is plus or minus $\id$.
In chapter 4 we defined an Abelian magnetic potential $A_{\rm mag}$
and field $F_{\rm mag}=d A_{\rm mag}$.
Now we wish to argue that the defects act as a source for the magnetic
field
$F_{\rm mag}$.
Moreover, we show that \textit{ in the absence of walls}\footnote{
We can formally define the absence of walls as follows.
Consider the extension of the defect manifold to $\R^3$, ie.
$\tilde \CD=\{\vx\in\R^3|\CP(\vx)=\pm\id\}$. There are no walls
if $\tilde\CD^c=\R^3\setminus \tilde \CD$ is connected.}
the total magnetic charge of the ${\cal P}=\id$
defects is quantised and is proportional to the instanton number
$q$.
The magnetic charge of the ${\cal P}=-\id$ defects is minus that of the
${\cal P}=\id$ defects so that the \sl total \rm
magnetic charge is zero.
This differs from the $\R^4$ case, where one only has magnetic
charge neutrality if one includes ``charges at infinity''.
In order to establish these results it is convenient to introduce a
static ``Higgs'' field $\phi(\vec{x})$ via
\bg
{\cal P}(\vx)=e^{i\phi(\vx)} ,\qquad \phi = \phi_j \sigma_j . \label{higgs0} 
\eg
Of course $\phi(\vx)$ is not globally defined.
Let ${\cal S}$ be a closed surface surrounding a ${\cal P}=\id$ defect.
We further \textit{ assume that ${\cal S}$ neither contains nor intersects
any other defects. }
For example, in fig. \ref{winding} an (extended) monopole defect 
 is surrounded by a
$2$-sphere
and  closed loop defects are surrounded by  2-tori.
 Now $\phi(\vx)$ may be smoothly defined on
${\cal S}$ and on the interior of ${\cal S}$ such that
the Higgs field is zero on the defect.
On ${\cal S}$ itself $\phi$ is non-vanishing and hence can be normalised.
The normalised field $\hat \phi = \phi / |\phi|$
takes its values in $S^2$ and defines a map
${\cal S}\rightarrow S^2$. The winding number of this map is \cite{novikov}

\eqnl{
n(\CS)={1\ov 16\pi i}\oint_{\CS} \Tr \left( \hat\phi\,
\d \hat \phi \wedge \d \hat \phi\right) .
}{magnwind3}
The magnetic flux through ${\cal S}$ is defined by
\eqnl{
\Phi(\CS)=\oint_{\CS} F_{\rm mag}.}{magnwind1}
In Appendix B we show that
\bg 
\Tr \left( \hat\phi\,
\d \hat \phi \wedge \d \hat \phi\right) =
8 i F_{\rm mag}
\eg
and hence 
\bg
n({\cal S})=\frac{1}{2\pi}\Phi({\cal S}).
\eg
That is the magnetic charge of the defect is proportional to the
winding number of the Higgs field
$\hat\phi:{\cal S}\rightarrow S^2$.
Hence it is quantised. Actually, if $\CS$ is a two-sphere
surrounding a magnetic monopole then $A_{\rm mag}\vert_\CS$ 
may be viewed as the Abelian gauge potential of
the Schwinger model on $S^2$ \cite{waia}, if $\CS$
is a two-torus as the gauge potential of the Schwinger model
on $T^2$ \cite{sachs}. The quantised flux $\Phi$ is just
the quantised instanton number of the Schwinger model on
$S^2$ or $T^2$.

We now look at the relation between the winding numbers of the
defects (and hence the magnetic charges) and the instanton number $q$.
We again assume that we only have no walls.
In this case the Higgs field, $\phi(\vx)$, can be assumed to be smooth
throughout
$T^3$ \sl except \rm at the the ${\cal P}=-\id$ defects, where it is 
ill defined.
At the ${\cal P}=\id$ defects the Higgs field is zero.
In Appendix B we derive the following relation between
$\hbox{Tr}\left(({\cal P}^{-1}d{\cal P})^3\right)$
and the Higgs field $\phi$
\bg
\label{niceformula}
2i \hbox{Tr}({\cal P}^{-1}d{\cal P})^3
= 3\d \Big[ \big( |\phi|-\hbox{${1\over 2}$}\sin(2 |\phi|)\big)
\Tr \big( \hat\phi \,\d\hat\phi\wedge d\hat\phi \big)\Big],
\eg
$\hat \phi$ being the normalised Higgs field.
Note that the right hand side of (\ref{niceformula})
is ill-defined both where ${\cal P}=\id$
and ${\cal P}=-\id$.
Using the instanton number formula
\refs{polloopindex} and \refs{niceformula} we can write the
topological charge as a functional of the Higgs field
\bg
q = \frac{1}{16 \pi^2 i} 
\int_{\CD^c} \d  \Big[\big( |\phi|-\hbox{${1\ov 2}$}\sin(2 |\phi|)\big)
\Tr \big(\hat\phi\, \d\hat\phi\wedge\hat d\phi \big)\Big].
\eg
Before we can apply Stokes theorem we must exclude closed sets
(with infinitesimal volume in $\CD^c$)
surrounding both ${\cal P}= \id$ and ${\cal P}=-\id$ defects.
Thus we have
\bg\label{charge}
q=\frac{1}{16 \pi^2 i} 
\sum_i
\oint_{{\cal S}_i}
 \big(|\phi|-\hbox{${1\ov 2}$}\sin(2 |\phi|\big)
\Tr \big( \hat\phi\, \d\hat\phi\wedge\hat\phi \big) .
\eg
where the ${\cal S}_i$ are surfaces 
surrounding the defects. 
Note that the factor $|\phi|-\hbox{${1\over2}$}\sin(2 |\phi|) $ 
behaves very differently
in the neighbourhoods of the two kinds of defects.
Near a defect with $\CP=\id$ the Higgs field 
tends to zero and the factor vanishes as $\sim |\phi|^3$.
Since  
$\oint_{\CS} \Tr( \hat\phi \, d\hat\phi\wedge\hat\phi )$ 
stays finite if we approach such a defect
the integrals in \refs{charge} vanish when the surrounding surfaces
approach defects with $\CP=\id$.
On the other hand, in the neighbourhood of the $\CP=-\id$ defects 
we have $|\phi|\sim\pi$ so that
$\big(|\phi|-\sin( 2 |\phi|)/2\big)\sim\pi$, from which follows that
\bg
q = \frac{1}{16\pi i}
\sum_{\hbox{\footnotesize{$\CP=-\id$ defects}}}
\oint_{{\cal S}_i}
\Tr\big( \hat\phi\,\d\hat\phi\wedge \d\hat \phi \big).
\eg
At this point it is convenient to define an alternative Higgs field
$\phi_{alt}(\vx)$ through
\eqnn{
\CP(\vx)=-\exp\left[ i\phi_{alt}(\vx) \right],}
where now $\phi_{alt}(\vx)$ is smooth and zero at ${\CP=-\id}$, but
ill defined at the $\CP=\id$ defects.
In the absence of walls we have that both $|\phi(\vx)|$
and $|\phi_{alt}(\vx)|$ are in the interval
$[0,\pi ) $. In ${\cal D}^c$ one has the following relations
between the two Higgs fields
\bg
\label{relation-phi-alt}
|\phi|=\pi-|\phi_{alt}|,\quad
\hat\phi_{alt}=-\hat\phi.
\eg
Using this we see that  the topological charge is proportional
to the sum of winding numbers of $\phi_{alt}$ around the ${\cal P}=-\id$
defects
\bg
q = - \frac{1}{16 \pi i}
\sum_{\hbox{\footnotesize{$\CP=-\id$ defects}}}\oint_{{\cal S}_i}
\Tr \big( \hat\phi_{alt}\, \d\hat\phi_{alt}\wedge \d\hat
\phi_{alt} \big).
\eg
The orientation of integration in this equation 
is such that the normal vector on the surface 
$ {\CS_i} $ points inside the surface (onto the monopole).  
But in equation \refs{magnwind3} the orientation of integration is opposite.
Therefore we conclude
\bg
q = \sum_{\hbox{\footnotesize{$\CP=-\id$ defects}}} n_{alt}({\cal S}_i).
\eg
Hence the instanton number is the sum over the winding numbers 
of the Higgs field $ \phi_{alt} $ at $ \CP = - \id $ defects. 
Taking into account equation \refs{relation-phi-alt} we obtain 
\bg
\Tr \left( \hat\phi_{alt}\, \d\hat\phi_{alt}\wedge \d\hat
\phi_{alt} \right) = - 8 i F_{\rm mag} .
\eg
Thus a $ \CP = - \id $ defect with winding number $ n_{alt} $ 
has magnetic charge $ - n_{alt} $.
Similar considerations yield that the instanton number $q$ is given by the 
sum over winding numbers of the Higgs field $ \phi $ at $ \CP = \id $
defects, or equivalently, by the sum of  all monopole charges 
at $ \CP = \id $ defects.
The relation between the instanton number and the magnetic charges of
\textit{ pointlike monopoles } on $\R^4$ has already been
obtained by Reinhardt \cite{reinhardt}.

For a non-zero flux the magnetic potential 
$A_{\rm mag}$ must necessarily be singular somewhere on $\CS$,
else the flux $\oint dA_{\rm mag}$ would vanish. 
As is well-known from the Dirac monopole, we may assume
that $A_{\rm mag}$ is regular on $\CS$ with one point
removed. Since this holds true for any $\CS\subset\CD^c$ surrounding
a charged defect we must attach a string
to each such defect on which $A_{\rm mag}$ is singular.
By definition, wherever the magnetic potential is singular
the diagonalisation matrix $W(\vx)$ is singular.

\begin{figure}[ht]
\begin{minipage}[ht]{14cm}
\centerline{\epsfysize=8 cm\epsffile{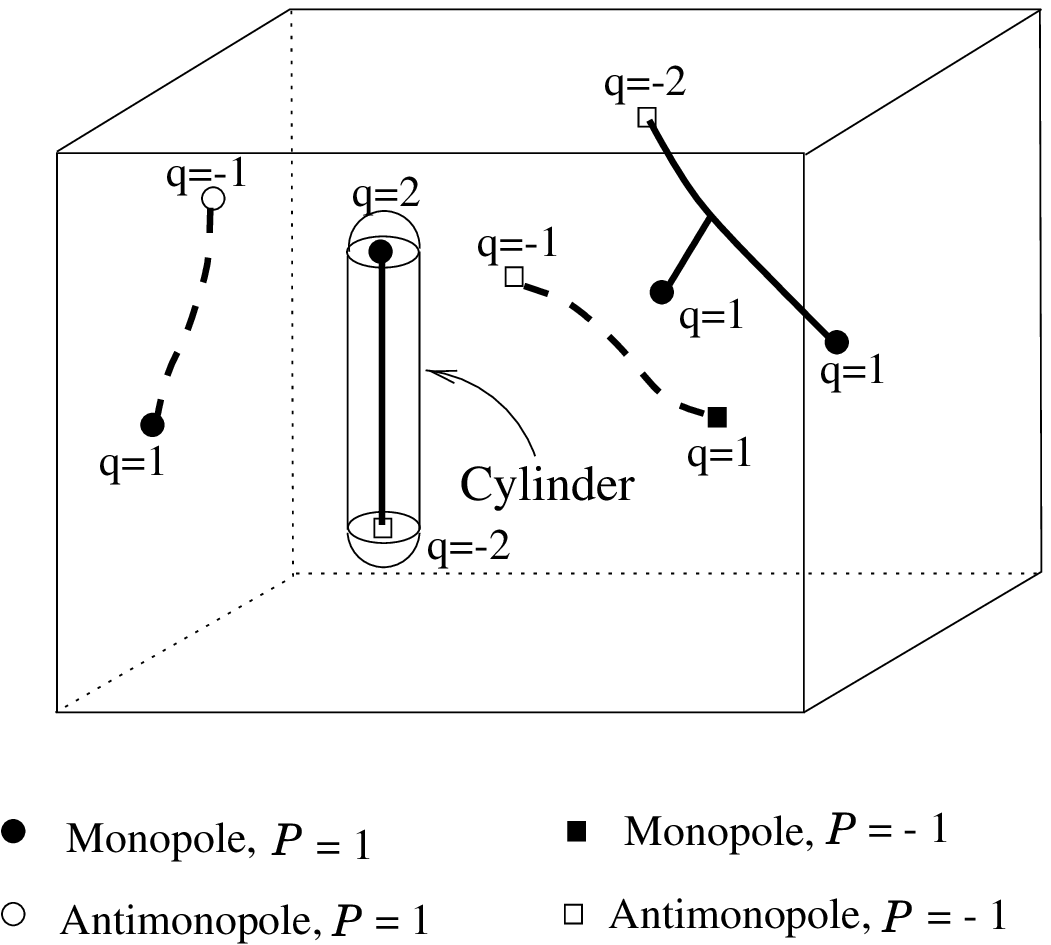}}
\caption{\label{monostrings}\textsl{The smooth diagonalisation
of $\CP(\vx)$ fails on strings connecting monopoles and
anti-monopoles of opposite charges. Shown is a string-network
in the sector with instanton number $q=4$}}
\end{minipage}
\end{figure}

Let us now consider a $S^2\subset \CD^c$ surrounding 
a monopole-antimonopole pair. On such a sphere the Higgs field 
has no winding and can smoothly be diagonalised.
This means that the strings on which $W(\vx)$ (and $A_{\rm mag}$)
is singular start and end at defects with  
opposite magnetic charge.  Outside of these strings it 
is possible to choose $W$ smooth.
A possible distribution of
monopoles connected by strings is shown in fig.\ref{monostrings}.
The string positions are gauge dependent. But they must
start and end at (anti)monopoles whose positions are
gauge invariant.
There is some freedom regarding which defects are connected to each
other with Dirac strings.
Suppose we have a Dirac string emanating from a $\CP=\id$ defect.
Then this string may be connected to \textit{ either }
a $\CP=\id$ or $\CP=-\id$ defect with the opposite charge.
We have shown that the instanton number, $q$,
is proportional to the total magnetic charge at ${\cal P}=\id$ defects.
We can restate this result in terms of the Dirac strings as follows;
$q$ is proportional to the number of Dirac strings
joining $\CP=\id$ and $\CP=-\id$ defects\footnote{Dirac 
strings joining $\CP=\id$ anti-monopoles to $\CP=-\id$ monopoles
count with a relative minus sign to strings joining $\CP=\id$
monopoles
and $\CP=-\id$ anti-monopoles.}.
In fact, it is possible to rewrite the instanton number
formula \refs{polloopindex}
so that the contribution of the strings is
transparent without introducing Higgs fields.
This calculation is given in Appendix B.

To gain further insight we investigate $\CP$ in
the vicinity of a point defect (see also \cite{griesshammer}).
For that we follow the eigenvalues along a closed path
from $p$ to $p$ (see fig.\ref{defects}a) passing through 
a monopole\footnote{a $SU(2)$-monopole 
with $\CP=\id$}.
We may slightly deform this path so that it  misses
the monopole. On the deformed path the eigenvalues
of $\CP$ are nowhere degenerate and thus $H$ in
\refs{diagonalization} must be periodic (see above). Returning
to the undeformed path through the monopole we conclude,
that at the monopole, where both eigenvalues are $1$,
the two eigenvalues are reflected at $\lam=1$
as shown in fig.\ref{defects}a.
The spectral flow is continuous but not differentiable.
Since $\CP(\vx)$ is smooth, the diagonalising $W(\vx)$
in \refs{diagonalization} must be singular to
compensate for the non-smoothness of $D(\vx)$.
As in the case without defects, 
$W(\vx)$ is not necessarily periodic and obeys \refs{hilfs1}.

To summarise, in the presence of
monopole and loop  defects we can still perform the gauge
transformation \refs{gauge-trf-1}. But it is
necessarily singular on strings connecting the defects.
The gauge transformed $A_0(\vx)$ is periodic, but singular 
on these strings. After gauge fixing the transition functions
can be chosen as in \refs{simpletrans}.
The instanton number is related to the magnetic charges of the defects
in a very simple way.

In cases where domain walls extend over the whole torus
the situation is analogous to the one in
$2$-dimensional gauge theories. We cannot
avoid defects when going from one ``face''
of the torus to the opposite one. There is no obstruction 
for the eigenvalues to vary smoothly along a
path crossing the wall. Neither $H$ nor $W$
are singular at the wall (see fig.\ref{defects}).
But the eigenvalues of $H$ may wind or interchange if
we move from $\vx$ to $\vx+b_i$ and thus $D(\vx)$ may be
periodic only up to a permutation of its entries, i.e.
\eqnn{
H(\vx+b_i)=\CW_i H(\vx)\CW_i^{-1}+\hat H_i,}
where the $\CW_i$ are Weyl-reflections and 
the $\hat H_i\in\CL$, with $\CL$ being the discrete
lattice defined in \refs{discrete}.
Now we can only prove, that the gauge transformed
transition functions have the form
\eqnn{
U_0=\id, \quad U_i=e^{-2\pi i\xi_0\hat H_i}R_i(\vx)\CW_i,}
where again the $R_i$ are in the Abelian subgroup
and fulfil the cocycle condition (\ref{cocycle}).
This time dependence in the transition functions is reminiscent
of the $T^2$ analysis \cite{mpw}.
 The gauge fixed
$A_0$ has the form
\eqnl{
A_0(\vx)=a_0(\vx)+{2\pi\ov L_0}\sum_{i=1}^3\hat H_i\xi_i,
\mtxt{where} a_0(\vx+b_i)=\CW_i a_0(\vx)\CW_i^{-1}.}{form}
As in the case without defects there are residual
 gauge transformations (see Appendix C).

\mysection{Conclusions and Outlook}

In this paper we have considered the gauge-fixing of Yang-Mills theory on
the four torus for arbitrary instanton sectors. 
Of course the choice of gauge fixing does not affect physics, but
an appropriate  gauge-fixing
may considerably simplify the mathematical problem
of computing or approximating functional integrals.
Motivated by their success in two dimensions we adopted
an extended Abelian projection on the four torus.  
Here we require $A_0$ to be Abelian, time independent and the 
spatial components $A_i$ of the gauge potential
to be periodic in time. 

One difference with the $T^2$  problem 
 is that we have the added complication of instanton sectors.
In two dimensions one may assume that before gauge fixing
the gauge potential $A_\mu$ is completely periodic,
ie. the transition functions are trivial $U_\mu=\id$.
In four dimensions, we may only assume this if we are in the
zero instanton sector (ie. the topological charge $q$ is zero).
We have argued that for $q\neq 0$ it is convenient to work with
a new set of non-Abelian transition functions.
With these transition functions the path ordered exponential,
$\CP(\vec{x})$, which
is central to the gauge fixing is completely periodic, even though
of course the gauge field itself is non-periodic.
Moreover, these transition functions include the odd instanton sectors
of $SU(2)$. To our knowledge, smooth untwisted transition functions for this
case have not been given before.

The most significant break with the two dimensional treatment
is the presence of  unavoidable singularities in the final gauge
fixed potential \cite{tHooft}.
These singularities are due to ambiguities in the diagonalisation
of $\CP(\vec{x})$ where the eigenvalues of $\CP(\vec{x})$ are degenerate
(for $SU(2)$ this degeneracy occurs where $\CP(\vec{x})=\pm\id$).
There is a close analogy between these defects (ie. points, loops
or surfaces where $\CP(\vec{x})$ is degenerate) and magnetic
charges in Yang-Mills-Higgs theories \cite{thooftpolyakov}.
We have presented a detailed analysis of the special case where one
only has point and loop defects, which can be interpreted as 
magnetic monopoles and magnetised loops.
The gauge fixed potential is smooth everywhere except for ``Dirac
strings'' joining monopole (loop) pairs.
The instanton number, $q$ is simply the number of magnetic charges
at the $\CP=\id$ defects.

>From one viewpoint  the existence of these magnetic defects
imply that our attempt to generalise the two dimensional fixing
has failed. We take the opposite view. It is a long standing
conjecture that confinement of color is produced
by dual superconductivity (of type II) of the QCD vacuum \cite{mandel}. 
Indeed, lattice calculations \cite{lattice2} indicate that
magnetic monopoles (or loops?) are the dominant infrared degrees of
freedom, at least in the maximal Abelian gauge and the Polyakov
gauge.

There is a long way from the picture of condensed magnetic
monopoles to real $QCD$. At present there is no analytic
proof of the existence of the condensate of monopoles. However,
in those theories where we  understand confinement,
the latter is due to the condensation of monopoles; these
examples are compact $QED$ \cite{poly} and supersymmetric
Yang-Mills theories \cite{seiberg}. The balancing of the
energy and the entropy of monopoles (and/or loops) may
explain the occurrence of the deconfinement transition
in $QCD$. At low temperatures we expect a condensation
of monopoles with $\CP=\id$ and of monopoles with $\CP=-\id$.
In the broken high temperature phase, where $\langle \Tr\CP\rangle\sim
\pm 2$, we do not expect long monopole loops but rather
a dipole gas of monopole-antimonopole pairs, both with
$\CP=\id$ (or both with $\CP=-\id$).

Of course the treatment given here has been purely classical. The next
step would be to study the path integral within this gauge fixing.
At this point one would need a suitable approximation
\cite{lenz}.
With a view to investigating the confinement of static quarks
it would be interesting to consider whether in any regime
the monopoles and Dirac strings  play a dominant role in the path integral.

\section*{Acknowledgements}

We are grateful to O. Jahn, F. Lenz,
G. Rudolph  and P. van Baal for helpful discussions.

\appendix 
\section*{Appendices}
\mysection{Theta Functions}

Our conventions with respect to theta functions are the same
as ref. \cite{tata}.
We work with the Jacobi theta function with characteristics
\bg
\theta\btensor{[}{c}
a \\ b \etensor{]}(z,i\tau)=
\sum_{n\in Z}
e^{-\pi\tau(n+a)^2+2\pi i(n+a)(z+b)}.
\eg
This function has the following periodicity properties 
\bg\theta\btensor{[}{c}
a +m\\ b+n \etensor{]}(z,i\tau)=
e^{2\pi i n a}\theta\btensor{[}{c}
a \\ b \etensor{]}(z,i\tau),\quad m,n\in Z\eg
and has zeros where $z=(a+n+\frac{1}{2})\tau+(b+m+\frac{1}{2}), \ n,m \in Z$.

\mysection{Technical Results}

In this appendix we derive some of the technical results quoted in
chapter 5.

\subsection{Magnetic Charges and Higgs winding numbers}

To relate this winding number to the magnetic
flux we parameterise the normalised Higgs field as
\eqnl{\hat\phi=\big(\sin\theta\cos\varphi,\sin\theta
\sin\varphi,\cos\theta\big),\qquad
\phi=\vert \phi\vert \hat\phi_i\sigma_i,}{higgs}
where the angles $\theta,\varphi$ are functions on $\CS$. 
The corresponding $\CP=\exp(i\phi)$ is diagonalised by
\cite{KR}
\eqnn{W=\exp(-i{\varphi\ov 2}\sigma_3)\exp(-i{\theta\ov 2}\sigma_2)
\exp(i\lam \sigma_3)\mtxt{and}
D=\exp(i\vert\phi\vert \sigma_3).}
The magnetic potential \refs{magnpot} is 
$A_{\rm mag}=d\lam-\ha \cos\theta d\varphi$,
and the corresponding  field strength is
\eqnn{
F_{\rm mag}=\ha\sin\theta \d\theta\wedge \d\varphi.}
On the other hand, taking the Higgs field \refs{higgs} we get
\bg
\Tr \left( \hat \phi \d \hat \phi \wedge \d \hat \phi \right)
= 
4 i \sin \theta \d \theta \wedge \d \varphi = 8 i F_{\rm mag} .
\eg 
Comparing with equations
\refs{magnwind3} and \refs{magnwind1} one readily obtains
\eqnl{
n(\CS)={1\ov 2\pi}\oint_{\CS}\sin \theta \;d\theta\wedge d\varphi=
{1\ov 2\pi}\Phi(\CS).}{magnwind4}

\subsection{ Derivation of equation \refs{niceformula}}

Now we will relate $ \Tr \left( \left( \CP^{-1} \d \CP \right)^3 \right) $
to the Higgs field $ \phi = \phi_\alpha \sigma^\alpha $. 
With the notation $ | \phi | = \sqrt{\Tr\phi^2 / 2}$ and 
$ \hat \phi := \phi / | \phi | $ we have $
{\cal P} = \exp \left( i \phi \right) = \cos( |\phi| ) 
+ i \sin( |\phi| ) \hat \phi
$ and it follows that
\bg
\label{p3}
( {\cal P}^{-1} \d {\cal P} )^3 &  = & 
- \sin^4| \phi |\hat \phi \d \hat \phi \wedge \d \hat \phi \wedge 
  \d \hat \phi 
- i \sin^3| \phi |\cos| \phi |\d \hat \phi \wedge \d \hat \phi \wedge 
  \d \hat \phi \\
&&\di - 3 i \sin^2| \phi |\hat \phi \d \hat \phi \wedge \d \hat \phi \wedge 
    \d| \phi |.
\eg
Under the trace the first two terms on the right hand side of \refs{p3} 
drop out. Hence we have
\begin{eqnarray}
\nonumber
\Tr \left( ({\cal P}^{-1} \d {\cal P} )^3 \right) & = & 
- 3 i \sin^2| \phi |\d| \phi |\wedge \Tr\left( 
\hat \phi \d \hat \phi \wedge \d \hat \phi \right) \\
\label{index-hat-phi}
& = &\d \left\{ \frac{3}{2 i} 
\left( |\phi| - \frac{\sin( 2 |\phi |)}{2} \right) 
\Tr \left( 
\hat \phi \d \hat \phi \wedge \d \hat \phi \right) \right\} .
\end{eqnarray}

\subsection{ Instanton number and Dirac Strings} 

We showed in the paper that in the presence of magnetic monopoles 
the diagonalization is not smoothly possible. 
The matrix $ W $ becomes singular on Dirac strings connecting monopoles 
with opposite magnetic charges. 
We shall argue that the strings on which $W$ is singular 
contribute to the instanton number $q$.
Setting $ \CP = W D W^{-1} $ one first observes for arbitrary gauge groups 
that
\eqnn{
\Tr \left( (\CP^{-1} \d \CP )^3 \right)= \d\CA,}
where the $2$-form $\CA$ is 
\eqnn{
\CA =-6\Tr \left(W^{-1}\d W \wedge D^{-1} \d D \right)+
3\Tr\left(W^{-1} \d W D^{-1}\wedge W^{-1} \d W D \right).}
Thus we can convert the integral in (\ref{polloopindex}) 
into a surface integral over the 
``boundary'' of the torus and over infinitesimal cylinders 
around the strings (see fig.\ref{monostrings}):
\eqnn{
q = \frac{1}{24 \pi^2} \int  
\Tr \left( (\CP^{-1} \d \CP )^3 \right) = 
q_{{\rm s}} + q_{{\rm b}},}
where the individual contributions from the strings and
boundary of the torus read
\eqnn{
q_{\rm s}= \frac{1}{24 \pi^2} \sum_{\rm strings} 
\;\intl_{\rm cyl.} \CA \;\mtxt{and}\;
q_{\rm b}= \frac{1}{24 \pi^2} \sum_{i=1}^3 
\Big(\intl_{x_i=0} \big( \CA(x+b_i )-\CA( x )\big)\Big).
}
If only monopoles are present, then $ H(\vx)$ is periodic 
and $ W(\vx+b_i)=W(\vx)R_i(\vx)$, where the $R_i$ are
abelian and satisfy the cocycle conditions \refs{cocycle-II}.
After some algebra we obtain 
\eqnn{
\CA(x+b_i) - \CA (x) = -
6 \Tr \big( R_i (x )^{-1} \d R_i(x) \wedge D(x)^{-1} \d D(x) \big).}
We parametrize the diagonal matrix as
\bg
D=e^{i \alpha \sigma_3} \mtxt{so that}
D^{-1} \d D = i \d \alpha \sigma_3
\label{parametrization}
\eg
and arrive at
\begin{eqnarray}
q_{\rm b} & = & 
{1\ov 4i\pi^2}
\sum_{i=1}^3 \int_{x_i = 0 } 
\d \Big[\al \Tr (R_i^{-1} \d R_i \sigma_3)\Big] \\
\nonumber
& = & \frac{1}{4 i \pi^2} \sum_{i,j} \int_{x_i = x_j = 0}
\varepsilon_{i j k} \alpha \Tr \left( \left(
R_i ( x + b_j )^{-1} \partial_k R_i ( x + b_j ) - 
R_i ( x )^{-1} \partial_k R_i ( x ) \right) i \sigma_3 \right) d x^k \quad .
\end{eqnarray}
Differentiating equation (\ref{cocycle-II}) one sees that 
the trace term is symmetric in $ i $ and $ j $. Therefore we 
conclude $ q_{\rm b} = 0 $ in accordance with the fact that the 
spatial transition
functions are simply given by the functions $ R_i $, see equation 
\refs{transition-fct-2D}.

The strings do contribute to the instanton number. 
We consider a Dirac string connecting monopoles. 
We argue that this string contributes to the instanton number the sum 
of monopole charges
of $ \CP = \id $ monopoles attached to the string.  
Using the parametrisation \refs{parametrization} we can write
\bg
\CA & = & - 12 \d \alpha \wedge A_{\rm mag} 
+ 12 \sin \alpha \cos \alpha F_{\rm mag} \\
& = & - 12 \d ( \alpha + \sin \alpha \cos \alpha ) \wedge A_{\rm mag} 
+ 12 \d ( \sin \alpha \cos \alpha A_{\rm mag}) .
\eg
Integrating over a closed surface $ \CS $ surrounding the string 
the contribution from the second term  vanishes. 
Now we choose a $ \CP = \id $ monopole and the Dirac string emanating from it. 
We introduce coordinates $ ( z , \varphi ) $ on $ \CS $ 
such that $ \alpha $ is independent of $ \varphi $. 
The 
contribution of the string to the instanton number reads
\bg
\frac{1}{24 \pi^2} \oint_{\CS} \CA = - \frac{1}{2 \pi^2}
 \int \d z \frac{\partial}{\partial z} 
\left( \alpha + \sin \alpha \cos \alpha \right) 
\int \d \varphi A_{{\rm mag} \, \varphi} .
\eg
The integral $ \int \d \varphi A_{{\rm mag} \, \varphi} $ is up to the sign 
given by the magnetic flux through the Dirac string, ie. 
it is $ - 2 \pi $ times the 
magnetic charge of the $ \CP = \id $ monopole.
Therefore the contribution of the string to the instanton number is given by
$ ( 1 / \pi )  \Delta ( \alpha + \sin \alpha \cos \alpha ) $.
Hence, if the string ends at a $ \CP = - \id $ monopole then it contributes 
$ 1 \,\, (\Delta \alpha = \pi ) $ and, if it ends at a 
$ \CP = \id $ monopole $ (\Delta \alpha = 0 ) $, 
it will not contribute. 
The generalisation to arbitrary strings is straightforward.

\mysection{Residual Gauge Fixing}

After the gauge fixing procedure described in sections 4 and 5,
$A_0$ is independent of time and restricted to the Cartan subalgebra.
Furthermore, the transition functions become abelian (upto an element
of the Weyl group if one has wall defects).
However, the gauge is not fixed completely, since one must
fix the residual gauge freedom related to gauge transformations
which preserve the properties of $A_0$ mentioned above.
If we have no walls we may assume that the transition functions have
the standard form (\ref{simpletrans}).
Thus we only consider residual gauge transformations which
do not change the transition functions.
One may regard this fixing of the transition functions as the first
part of our residual gauge fixing.
Let us first consider the case considered in chapter 4, where one
has no defects.

\subsection{ No defects}

Here we assume that the defect manifold ${\cal D}$ is empty, in which
case ${\cal P}(\vx)$ is smoothly diagonalisable.
We may also assume that $H(\vx)$ is smoothly restricted to the first
Weyl chamber.
After the first part of the gauge fixing given in section four
the transition functions are abelian.
The  residual gauge transformations are 
\begin{equation}
\label{resgauge1}
V(x)= \exp\left\{ 2\pi i
\left(H_{per}(\vx )+H_i\frac{x_i}{L_i}\right)\right\},
\end{equation}
where all $H$'s are in the Cartan subalgebra,  $H_{per}$
is periodic in all spatial directions, and $H_{i}\in {\cal L}$.
Clearly these residual gauge transformations have no effect on $A_0$.
Accordingly, to fix the gauge we must impose constraints on the
\textit{ spatial }
components of the gauge field.
Of course, $A_i(x)$ depends on time, whereas the residual gauge
transformations
under consideration are time-independent.
Thus we could impose constraints on $A_i$ for some fixed time, say
$x_0=0$.
Alternatively, if we wish to treat all times on an equal footing we
can consider the time-averaged object
\bg
\tilde A_i(\vx)=\frac{1}{L_0}\int^{L_0}_0dx_0 A_i(x_0,\vx).\eg

Using the result that the transition functions are Abelian 
\em{ after } \rm gauge-fixing, the Cartan part of $\tilde A_i(\vx)$
(or $A_i(x_0=0,\vx)$) may be decomposed into a periodic piece
$\tilde A_i^{c,per}$ and a contribution $\tilde A_i^{c,lin}$, 
which
is linear in the spatial coordinates.
If we  impose
\bg
{1\over {L_1 L_2 L_3}}\int_{T^3}
\tilde A_i^{c,per}d^3x  
\in \;{2\pi\ov L_i}\cdot {\cal
  H}/{{\cal L}},\quad
i=1,2,3,\eg
we  fix the gauge freedom with respect to the $H_i$.
Here ${\cal H}/{\cal L}$ is the torus obtained
by dividing the Cartan subalgebra ${\cal H}$
by the lattice ${\cal L}$.

If we demand the following relations, we can fix the residual gauge
freedom concerning $H_{per}$ upto an Abelian global gauge transformation 
\bg\label{first}
\tilde A^{c,per}_1 = h_1(x_2,x_3),\quad
\intl_0^{L_1}\tilde A^{c,per}_2 dx_1 =h_2(x_3),\quad
\intl_0^{L_1}\intl_0^{L_2}\tilde A^{c,per}_3dx_1dx_2=h_3,\eg
where the functions $h_i$ are Cartan subalgebra valued functions
of the relevant spatial coordinates.
An alternative to (\ref{first}) which is symmetric in the spatial
variables is simply the Coulomb type condition
\bg
\nabla\cdot \tilde {\bf A}^{c,per}(\vx)=0.\eg

\subsection{ Defects without walls}
 
We now assume that the defect manifold, ${\cal D}$, is non-empty.
Thus our gauge fixed potential is not well defined for $\vx\in{\cal
  D}$
and on the Dirac strings
 joining the defects.
While ${\cal D}$ is gauge invariant, the paths taken by the
Dirac strings are not.
However, one can only change the path of the Dirac strings
with a \em{ singular } \rm gauge transformation.
The residual gauge transformations considered in
the previous subsection were (implicitly) assumed to be smooth.  
Thus it is convenient to separate the residual gauge fixing into
two parts. Firstly, one fixes the location of the Dirac
strings (which can be viewed as a singular residual gauge fixing).
For example consider the case where one only has
point-like monopole defects.
Here one can take the Dirac strings to be straight lines all
meeting together in the centre of $T^3$.
Then one repeats the  residual gauge fixing of section C.1,
except that now 
one must exclude a closed set, $\cal G$, containing both the defect
manifold $\CD$
and the (by now fixed) Dirac strings from the relevant integrals in (\ref{first}).

\subsection{Walls}

If we allow for wall defects then we actually have a wider class of
residual gauge transformations
\begin{equation}
\label{resgauge2}
V(x)= {\cal W}\cdot \exp\left\{ 2\pi i
\left(H_{per}(\vx )+H_i\frac{x_i}{L_i}+H_0\frac{x_0}{L_0}
\right)\right\},
\end{equation}
where $H_0\in{\cal L}$, and ${\cal W}$ is an element of the Weyl
group \sl which commutes with all the transition functions.\rm
The reason for this extra freedom is that in the cases considered in
the previous subsections we could  assume that $H(\vx)$ is restricted to
the first Weyl chamber, and we only considered residual gauge
transformations
which respected this constraint.

\end{document}